\documentclass[ reprint,superscriptaddress,amsmath,amssymb,aps,showpacs, longbibliography]{revtex4-2}

\usepackage{comment}
\usepackage{enumitem} 
\usepackage{lgreek}

\usepackage{savesym}

\usepackage[utf8]{inputenc}
\usepackage[T1]{fontenc}
\usepackage{mathptmx}
\usepackage{etoolbox}

\savesymbol{amalg}
\savesymbol{corresponds}
\usepackage{mathabx}
\restoresymbol{BX}{amalg}
\restoresymbol{BX}{corresponds}

\usepackage{bibentry}
\newcommand{\ignore}[1]{}
\newcommand{\nobibentry}[1]{{\let\nocite\ignore\bibentry{#1}}}
\newcommand{\bibfnamefont}[1]{#1}
\newcommand{\bibnamefont}[1]{#1}

\usepackage{wasysym}
\usepackage{stmaryrd}
\usepackage{amsfonts}
\usepackage{xcolor}

\usepackage{graphicx}
\usepackage{dcolumn}
\usepackage{bm}
\usepackage{hyperref}

\begin{document}

\title[Three length scales colloidal gels: the clusters of clusters \textit{versus} the interpenetrating clusters approach]{Three length scales colloidal gels: the clusters of clusters \textit{versus} the interpenetrating clusters approach}

\author{Louis-Vincent Bouthier}
\email{louis.bouthier@etu.minesparis.psl.eu}
\affiliation{Groupe CFL, CEMEF, Mines Paris, PSL Research University, 1 Rue Claude Daunesse, 06904 Sophia Antipolis, France}
\author{Thomas Gibaud}
\email{thomas.gibaud@ens-lyon.fr}
\affiliation{ENSL, CNRS, Laboratoire de physique, F-69342 Lyon, France}

\date{\today}

\begin{abstract}
Typically, in quiescent conditions, attractive colloids at low volume fractions form fractal gels structured into two length scales: the colloidal and the fractal cluster scales.
However, when flow interfere with gelation colloidal fractal gels may display three distinct length scales (Dagès, et al., \emph{Soft Matter} 18, 6645 (2022)). Following those recent experimental investigations, we derive two models that account for the structure and the rheological properties of such atypical colloidal gels. The gel elasticity is inferred from scaling arguments and the structure is translated into scattering intensities following the global scattering functions approach proposed by Beaucage and typically measured in small angle X-ray scattering (SAXS).
In both models, we consider that the colloids condensate into fractal clusters. In the clusters of clusters model, the clusters form superagregates which then build the gel network. In the interpenetrating clusters model, the clusters interpenetrate one-another to form the gel network. Those two models are then used to analyse rheo-SAXS experiments carried out on carbon black gels formed through flow cessation.
\end{abstract}

\maketitle

\section{Introduction}

Colloidal gels are out of equilibrium soft solids composed of attractive brownian particles at low volume fraction dispersed in a fluid medium that self-assemble into space spanning network~\cite{trappe2001}. These viscoelastic materials are ubiquitous both in nature and in industrial applications as diverse as flow batteries, food products and cementitious materials \cite{Lu:2013,Ioannidou:2016,Parant:2017,Cao:2020,Macosko1994,Mewis2012,Wagner2021,Hengl2014,Ioannidou2016,Awad2012,Knorr2004,Chandrapala2012,gibaud2012}. Understanding the interplay between their mechanical and structural properties remains a challenge. 
In the limit of low volume fractions and strong attractions, colloids condensate into fractal gels. Under quiescent conditions, the gelation is driven by thermal motion and dynamical arrest. The aggregation mechanism are classified into diffusion-limited aggregation~(DLCA) or reaction-limited aggregation~(RLCA)~\cite{Weitz1984,Weitz1985,Jungblut2019}. 
Both scenario leads to a structuration of the gel into two length scales: the particles aggregate and form fractal clusters that fill the space~\cite{Weitz1984,Weitz1985}. The elasticity of RLCA and DLCA gels is well captured by $\phi$-power law models~\cite{Shih1990,Krall:1998}.

Unsurprisingly, given the fact that gels are out of equilibrium states, gelation under flowing conditions leads to structures that significantly differ from ones that are formed in quiescent conditions. Simulations indicate that the fluid flow play a strong role in defining the resulting structures~\cite{Jamali2020, Nabizadeh2021}.  Zaccone et al. \cite{zaccone2011} suggest that upon flow cessation, colloids condensate in sticky large non Brownian aggregates that are jammed. Experimentally, it is shown that flow may induce structural anisotropy~\cite{pignon1997, park2017, sudreau2022}, overaging~\cite{sudreau2022b}, heterogeneities~\cite{Koumakis2015}, microcracks~\cite{Gibaud2020a, Dages2021} or shear banding~\cite{Moller2008, divoux2016}.
In the presence of flow, attractive colloids at low volume fractions may also display a hierarchical structure composed of three distinct length scales. For instance, during fuel combustion, soot particles form and aggregate in submicrometer fractal clusters with a fractal dimension around $1.8$ which then aggregate in denser supramicrometer superaggregates with a fractal dimension around $2.6$~\cite{Kim2004,Sztucki2007}. Additionally, suspensions of carbon black particles dispersed in mineral oil subject to a flow cessation form gels which present three characteristic length scales related to fractal clusters of carbon black particles that interpenetrate one another~\cite{Dages2022}.

In section~\ref{sec:Experiment1}, we first present the measured mechanical and structural properties of a three length scales carbon black gels resulting from flow cessations~\cite{Dages2022}. We then discuss the implications of such results at a fundamental level and in terms of applications to motivate the models developed in the next section.
In the section~\ref{sec:Model}, we present the structure of two models that display three distinct length scales: the clusters of clusters model which describes gels structured in a network of superaggregates formed by clusters of particles~\cite{Bouthier2022b} and the interpenetration of clusters model which describes gels formed by a network of interpenetrating clusters of particles~\cite{Dages2022}. We then fit the small angle X-ray scattering data (SAXS) presented in the section~\ref{sec:Experiment1} with both models using a global scattering functions approach proposed by Beaucage.
In section~\ref{sec:rheo}, we derive the mechanical properties of the clusters of clusters model and the interpenetration of clusters model previously introduced. The derivation of such mechanical models is greatly inspired by the $\phi$-power law models initially built to describe classical colloidal gels with two distinct length scales, typically obtained in the DLCA or the RLCA context. The derivation yield an analytical expression of the gel elastic modulus. This is the main result of this paper. This analytical expression combined with the assessment of the  gel structure obtained by SAXS allows us to fit the gel elasticity $G'$ as function of the gel length scales, cluster fractal dimension, volume fraction, colloidal interactions, etc... 
The results of the fits indicate that the three length scale carbon black gels obtained through flow cessations~\cite{Dages2022} can be fitted by both models. We however identity two general trends: the gel elasticity originate from stretching bonds at all length scales and the gel structure become denser as the gel characteristic length scales increase.

\section{three length scale carbon black gels resulting from flow cessations}
\label{sec:Experiment1}

\begin{figure}
    \centering
    \includegraphics[width=1\columnwidth]{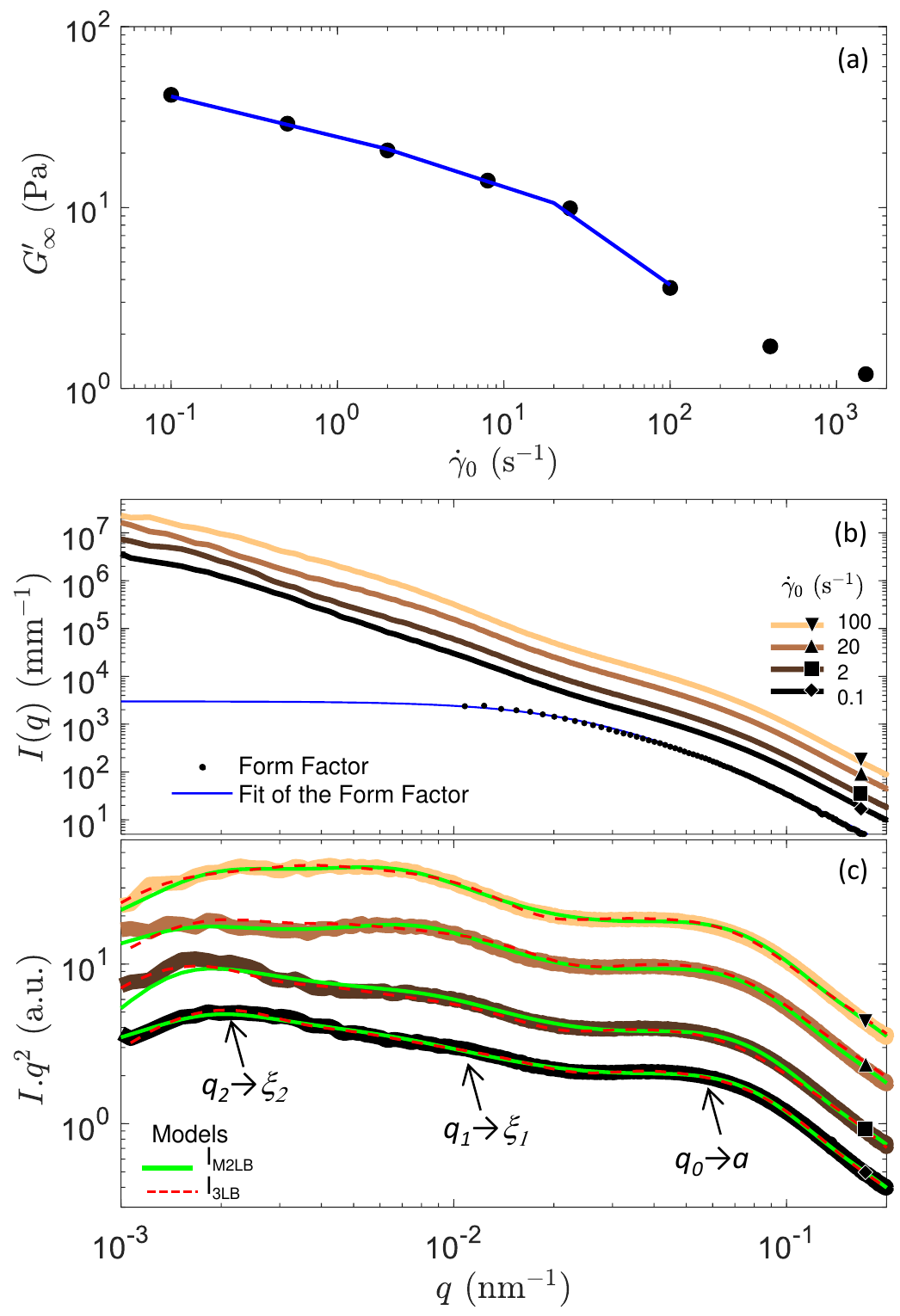}
    \caption{ Carbon black gels properties resulting from flow cessation. (a) Evolution of the gel elasticity $G'_\infty$  ($\bullet$) as a function of $\dot{\gamma}_0$  the shear rate intensity before flow cessation. the blue line represent the fit (a) through (d) listed in Tab.~\ref{tab:model}. The fits are indistinguishable. (b) Gel scattered intensity $I\left(q\right)$ as a function of the wave vector number $q$. The colors from light orange to black correspond to a decreasing $\dot{\gamma}_0$. The black dots correspond to the experimental form factor of the carbon black particles and the blue solid line is the fit of this form factor. $I(q)$ resulting from $\dot{\gamma}_0=0.1$~s$^{-1}$ is on absolute scale. The rest of the data is translated along the $y$-axis for better readability. (c) Kratky representation, $I\left(q\right)q^2$ versus $q$, of the data  shown in (b). The arrows point to the three bumps at $q_0$, $q_1$ and $q_2$, associated respectively to the length scales $a$, $\xi_1$ and $\xi_2$ The green lines correspond to the two level modified Beaucage model (Eq. (\ref{eq:Beaucage1})) and the red  dash lines correspond to three level beaucage model (Eq. (\ref{eq:Beaucage2})). Experimental data are extracted from ref.~\cite{Dages2022}.
    }
    \label{fig:Intensity}
\end{figure}

Carbon black particles are fractal carbonated colloids that result from the partial combustion of hydrocarbon oils \cite{lahaye1994,xi2006,Sztucki2007}. These particles are widely used in the industry for mechanical reinforcement or to enhance the electrical conductivity of plastic and rubber materials \cite{wang2018}. Dispersed in oil, the carbon black  particles are attractive and form gels at low volume fractions~\cite{trappe2000}.
Such gels present peculiar rheological properties~\cite{Gibaud2020b} including  rheopexy \cite{Ovarlez:2013,Helal:2016,Hipp:2019}, delayed yielding \cite{Gibaud:2010,Grenard:2014}, fatigue \cite{Gibaud:2016, Perge:2014} and rheo-acoustic properties~\cite{Gibaud2020a, Dages2021}. Here, we are interested in revisiting carbon black gels resulting from flow cessations~\cite{Dages2022}. The flow cessation protocol were carried out in carbon black dispersion (Vulcan PF) at $c=4~\%_w$ ($\phi=1.58$~\%) in mineral oil and is described in~\cite{Dages2022}. In brief, the protocol consist in (i) a rejuvenation step at a strain rate $\dot{\gamma}=1000~\mathrm{s}^{-1}$ during 60~s, (ii) a preshear step at $\dot{\gamma}=\dot{\gamma}_0$ during 20 to 200~s, (iii) a flow cessation induced by $\sigma=0~\mathrm{Pa}$ for 30~s, (iv) a rest period of 360~s during which the viscoelastic moduli are monitored and finally (v) a frequency sweep or a SAXS measurement. The gel mechanical and structural properties (Fig.~\ref{fig:Intensity}) as measured in step (v) of the protocol depend on the value of $\dot{\gamma}_0$, the shear rate carried out before flow cessation.

 Fig.~\ref{fig:Intensity}(a) displays the evolution of the elastic modulus $G'_\infty$ of the gel measured during a frequency sweep experiment in the low frequencies domain as a function of $\dot{\gamma}_0$ the shear intensity before flow cessation. The gel become stronger as $\dot{\gamma}_0$ decreases: its elasticity can be tuned by a factor 50, from $\sim 1$~Pa at high $\dot{\gamma}_0$ to $\sim 50$~Pa at low $\dot{\gamma}_0$.

 The structural properties of the gel as a function of $\dot{\gamma}_0$ are determined based on SAXS experiments. The scattered intensities $I(q)$ as a function of the wave number $q$ are displayed in Fig.~\ref{fig:Intensity}(b). The $I(q)$ are isotropic in the $q$-range tested. $I(q)$ show the usual decreasing trend due to the proportional relationship of the intensity with the number of scatterers in volume defined by the lenthscale $1/q$. 
{\color{black} More precisely, the intensity roughly decay as $1/q^2$ which is expected for fractal gels. We therefore turn to the  Kratky representation in Fig.~\ref{fig:Intensity}(c) and displays $Iq^2$ versus $q$. In this representation, a structure of fractal dimension 2 at all length scale is an horizontal line. Any deviation from this flat line are highlighted by the Kratky representation.}
 We observe three bumps at low, intermediate and high $q$ related to three length scale. Those bumps are particularly visible in the Kratky representation in Fig.~\ref{fig:Intensity}(c). The high-$q$ bump at $q_0$ is related to the radius $a$ of the carbon black particles and the two other peaks at intermediate-$q$ ($q_1$) and low-$q$ ($q_2$) correspond to two additional length scales $\xi_1$ and $\xi_2$ respectively.
This structure is atypical. Indeed, in gels driven solely by thermal agitation, $I(q)$ classically displays only two characteristic length scales: the particle size $a$ and the cluster size $\xi$ separated by a power-law regime, the exponent of which is related to the cluster fractal dimension \cite{courtens1987}.


Those rheology and SAXS results have strong implications. On the application level, such an interplay between shear and gelation is involved in numerous industrial processes, and especially in additive manufacturing  where shear coupled with 3D printing allows tuning the microstructure and the properties of the printed materials~\cite{Raney:2018}. On a fundamental level, as already well discussed in the literature, shear may interfere with the gelation pathway of particulate colloidal gels and lead to multiple metastable gels which properties, microstructure~\cite{Koumakis2015, das2022}, connectivity~\cite{Helal:2016} or yield stress~\cite{Ovarlez:2013} depend on the flow cessation protocol. Experiments displayed in Fig.~\ref{fig:Intensity} are  ideal for exploring the interplay between microstructure and mechanics: as the gel originates from the same dispersion, the colloid volume fraction and interactions remain unchanged, while variations in the gel mechanical properties only result from microstructural changes. 

\begin{figure*}
    \centering
    \includegraphics[width=1.7\columnwidth]{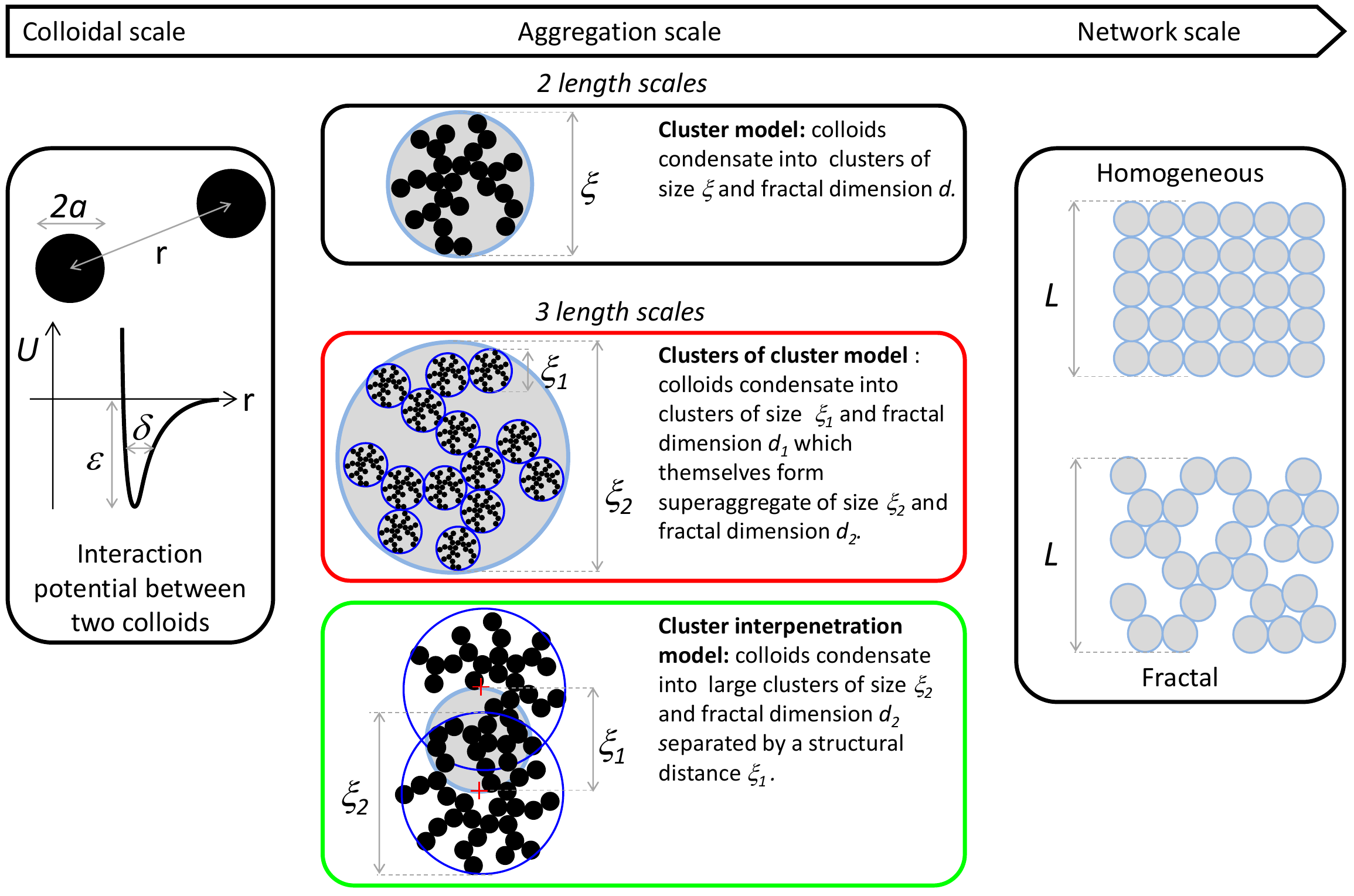}
    \caption{Sketch of the gel hierarchical structures. $\phi$-power law models which model the gel elasticity $G'$ are based at the particle scale on the colloid-colloidal interactions, at the meso scale on the aggregated structures and at the macroscopic scale on the network topology. The structural length $\xi_s$ defined at the aggregation scale sets the coarse grain scale for the network scale. In  the cluster $\phi$-power law model $\xi_s=\xi$, in the clusters of clusters $\phi$-power law model $\xi_s=\xi_2$ and in the interpenetration of clusters $\phi$-power law model $\xi_s=\xi_1$.}
    \label{fig:SketchMix}
\end{figure*}

{\color{black} In~\cite{Dages2022}, the three length scales were interpreted using the interpenetration cluster model in the weak-link regime assuming an homogeneous network. This analysis, as mentioned in~\cite{Dages2022}, has however some serious drawbacks that we intend to address in this paper.  First, there is a discrepancy between the local particle density $\rho$ obtained from the SAXS analysis and the average density $\bar \rho$ obtained from the carbon black concentration: $\rho > \bar \rho$. Second, using the interpenetration model in the weak-link regime for an homogeneous network, the elasticity resulting from the carbon black inter particle interaction $G_{\mathrm{cc}}$ is underestimated by orders of magnitudes. This suggest that the network in heterogeneous. Finally, the SAXS data can also be interpreted differently, in clusters of clusters ~\cite{Kim2004,Sztucki2007} rather than in interpenetrated clusters. In the next section, as it is currently lacking in the literature, we develop three-length-scales gel rheological models to address the above problems and to establish a quantitative relation between the structure and the elasticity of carbon black gels resulting from flow cessation. }

\section{Structure of the gels}
\label{sec:Model}

 We construct two three-length-scales gel models to fit the data presented in the previous section. The clusters of clusters $\phi$-power law model describe gels structured in a network of superaggregates formed by clusters of particles. The interpenetration of clusters $\phi$-power law model describe gels formed by a network of interpenetrating clusters of particles. The models structure are sketched in Fig.~\ref{fig:SketchMix}. In this section, we first present the model from a structural perspective then we fit and discuss the SAXS data.

\subsection{Structural models}

In the clusters of clusters $\phi$-power law model~\cite{Bouthier2022b}, we consider that colloidal particles of radius $a$ assemble into clusters of  dimension $\xi_1$ and fractal dimension $d_{1}$ which themselves aggregates into  super-aggregates of  dimension $\xi_2$ and fractal dimension $d_{2}$ to form the gel network as sketched in the Fig.~\ref{fig:SketchMix}. In Fig.~\ref{fig:Intensity}(c), the high-$q$ peak at $q_0$ corresponds to the carbon black particles of radius $a$, the intermediate-$q$ peak corresponds to the cluster size $\xi_1$ and the low-$q$ peak corresponds to the super-aggregates of size $\xi_2$. The fractal dimensions $d_1$ and $d_2$ are related to the slope of the scattering intensity $I(q)$ measured between the peaks ($q_0$, $q_1$) and ($q_1$, $q_2$) respectively.

The intensity spectrum $I(q)$ may then be fitted by a three-level Beaucage model\cite{Beaucage1995,Beaucage1996,Hammouda2010} through 

\begin{equation}
    I_\mathrm{3LB}\left(q\right)=I_2\left(q\right)+I_1\left(q\right)+I_a\left(q\right),
\end{equation}
with
\begin{equation}
\begin{array}{l}
\left\{
\begin{array}{ll}
I_2\left(q\right) & = G_2 \mathrm{e}^{\left(-\frac{q^2\xi_2^2}{3}\right)}+
    B_2  \mathrm{e}^{\left(-\frac{q^2\xi_1^2}{3}\right)}\mathrm{erf}\left(\frac{q\xi_2}{\sqrt{6}}\right)^{3d_2}q^{-d_2} \\
I_1\left(q\right) & = G_1 \mathrm{e}^{\left(-\frac{q^2\xi_1^2}{3}\right)}+B_1\mathrm{e}^{\left(-\frac{q^2a^2}{3}\right)}\mathrm{erf}\left(\frac{q\xi_1}{\sqrt{6}}\right)^{3d_1}q^{-d_1} \\
I_a\left(q\right) & = G_0\mathrm{e}^{\left(-\frac{q^2a^2}{3}\right)}+B_0\mathrm{erf}\left(\frac{qa}{\sqrt{6}}\right)^{3d_0}q^{-d_0}
\end{array}
\right.
\end{array}.
\label{eq:Beaucage2}
\end{equation}

The three-level Beaucage model sum the scattering contribution $I_2(q)$ of the superaggregates of size $\xi_2$ and fractal dimension $d_2$, the contribution $I_1(q)$ of the clusters of size $\xi_1$ and fractal dimension $d_1$ and the contribution $I_a(q)$ of the constituent colloidal particles of size $a$ and fractal dimension $d_0$.
The terms multiplied by $\left(G_i\right)_{i=0,1,2}$ correspond to the approximation of the Guinier regime when the wave vector number tends towards 0 which vanishes when the wave vector number goes above the inverse length scale designated by each level $a$, $\xi_1$ or $\xi_2$. The terms multiplied by $\left(B_i\right)_{i=0,1,2}$ correspond to the approximation of fractal behaviour with a power-law scaling in $q$ when its values are above the designated length scale and vanishing values relatively to the Guinier regime when $q$ tends to 0 with the combination of the error function and the power-law in $q$. The exponential terms associated with the terms multiplied by $B_1$ and $B_2$ are here to make the contribution vanish when the smaller length scale level is reached.

In the interpenetration of clusters $\phi$-power, we consider that colloidal particles of radius $a$ assemble into clusters of size $\xi_2$ and fractal dimension $d_2$. The cluster of size $\xi_2$ interpenetrate one another, such that their center to center distance $\xi_1$ is smaller than $\xi_2$, {\color{black} to form a network as sketched in the Fig.~\ref{fig:SketchMix}}.  In Fig.~\ref{fig:Intensity}(c), the high-$q$ peak at $q_0$ corresponds to the carbon black particles of radius $a$. The intermediate-$q$ peak is a structural peak and corresponds to $\xi_1$ the center to center distance between two adjacent clusters. The low-$q$ peak corresponds to the cluster of size $\xi_2$ and fractal dimension $d_2$. The fractal dimensions $d_2$ is related to the slope of the scattering intensity $I(q)$ measured between the peaks ($q_0$, $q_2$).

The intensity spectrum $I(q)$ may then be fitted by by a modified two level Beaucage model~\cite{Beaucage1995,Beaucage1996,Keshavarz2021} through 

\begin{equation}
    I_\mathrm{M2LB}\left(q\right)=I_2\left(q\right)S_1\left(q\right)+I_a\left(q\right)
\end{equation}
with
\begin{equation}
\begin{array}{l}
\left\{
\begin{array}{ll}
I_2(q) & = G_2 \mathrm{e}^{\left(-\frac{q^2 \xi_2^2}{3}\right)} + B_2 \mathrm{e}^{\left(-\frac{q^2 a^2}{3}\right)}\mathrm{erf}\left(\frac{q \xi_2}{\sqrt6}\right)^{-3d_2}   q^{-d_2} \\
S_1(q) & =1+C_1 \left( \left(\frac{q\xi_1}{2\pi}\right)^2 + \left(\frac{2\pi}{q\xi_1}\right)^2 \right)^{-1} \\
I_a(q) & =G_0 \mathrm{e}^{\left(-\frac{q^2 a^2}{3}\right)} + B_0\mathrm{erf}\left(\frac{q a}{\sqrt6}\right)^{-3d_0}q^{-d_0}
\end{array}
\right.
\end{array}.
\label{eq:Beaucage1}
\end{equation}

The two-level Beaucage model sums the scattering contribution $I_2(q)$ of the clusters of size $\xi_2$ and fractal dimension $d_2$ and the contribution $I_a(q)$ of the constituent colloidal particles of size $a$. To account for the clusters interpenetration, the cluster intensity $I_2(q)$ is multiplied by an ad-hoc inter-cluster structure factor $S_1(q)$ leading to an increase of scattering at intermediate $q$. $S_1(q)$ is function that peaks at $q_1=2\pi/\xi_1$ to a maximum value $1+C_1/2$ and that converges to 1 away from $q_1=2\pi/\xi_1$. Such a choice of $S_1(q)$ is simple but not completely satisfactory as it fails in the thermodynamic limit. Indeed, $S_1(q\rightarrow 0)=1$ whereas it should be proportional to the isothermal compressibility.

In the beaucage models, the length scales $a$, $\xi_1$ and $\xi_2$ reflect a high order moment of the aggregate log-normal size distribution in the Beaucage functions and are indicative of the higher limit of the size distribution of the aggregates~\cite{kammler2005,beaucage2012}.

\subsection{Discussion of the gel structure}

We now fit the SAXS data shown in Fig.~\ref{fig:Intensity}(b-c) using the structural models for the clusters of cluster model (Eq.~\ref{eq:Beaucage2}) and the interpenetration cluster model (Eq.~\ref{eq:Beaucage1}). As shown in Fig.~\ref{fig:Intensity}(c), the scattering intensities are nicely fitted using both models. The fits are carried out in log-scale on data re-sampled with 20 points per decades. 
. 

\begin{figure}
    \centering
    \includegraphics[width=1\columnwidth]{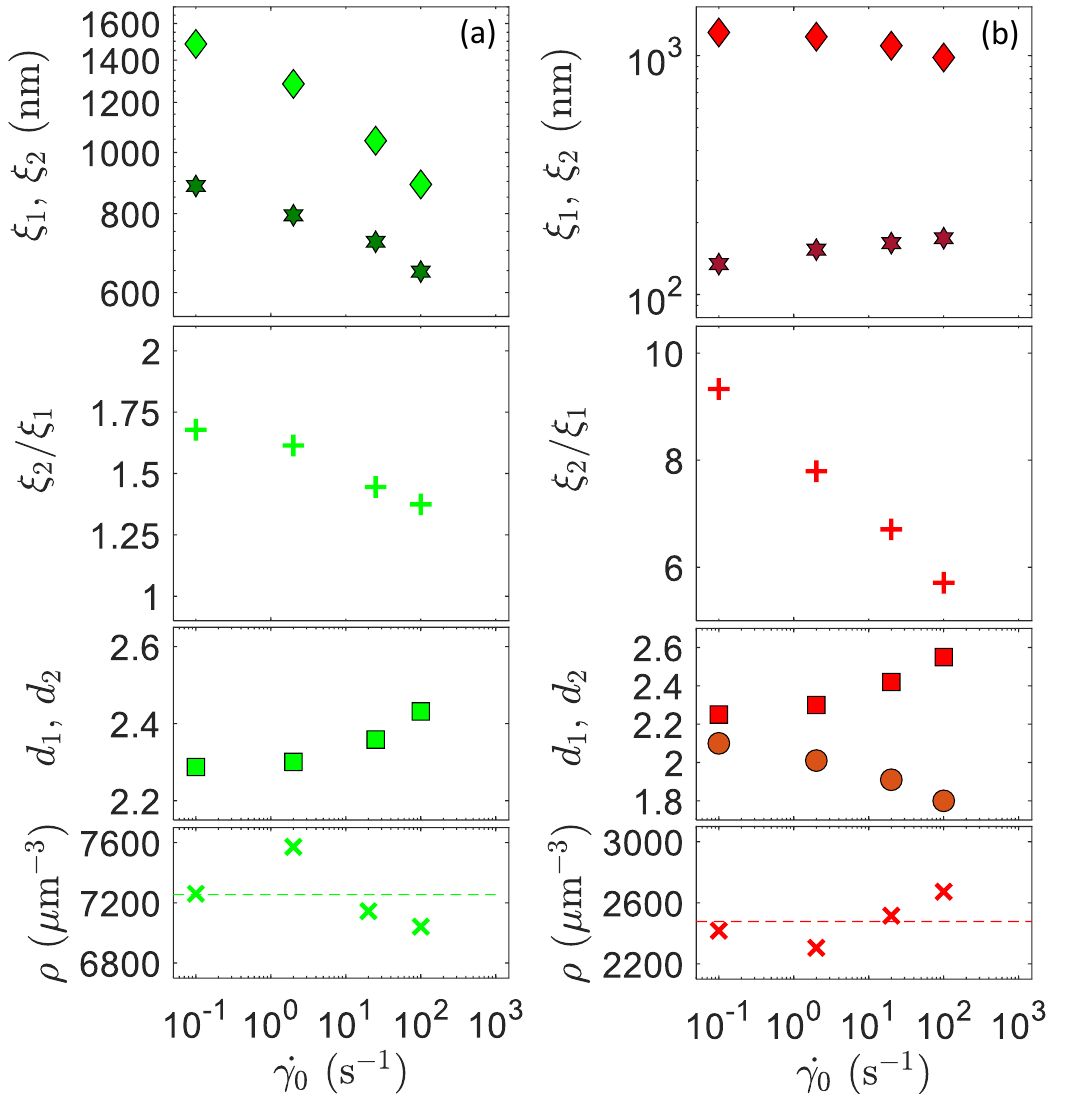}
    \caption{Parameters of the scattering models resulting from fitting the scattering intensity $I(q)$ in Fig.~\ref{fig:Intensity}, (a) using  the cluster interpenetration approach (Eq.~(\ref{eq:Beaucage1}), green symbols) and (b) using the clusters of the clusters approach (Eq.~(\ref{eq:Beaucage2}), red symbols). From top to bottom: $\xi_1$ ($\blacklozenge$), $\xi_2$ ($\bigstar$), $\xi_1 /\xi_2$ ($+$), $d_1$ ($\bullet$), $d_2$ ($\blacksquare$) and $\rho$ ($\times$) as a function of the applied shear rate $\dot{\gamma}_0$ before flow cessation. The carbon black radius $a$ is constant in all the fits and equal to $a=27$~nm.
    }
    \label{fig:Parameters}
\end{figure}

{\color{black} Let us first discuss the fit results obtained using the cluster interpenetration model. In Fig. \ref{fig:Parameters}a, as $\dot{\gamma}_0$ increases, we observe that the gel is composed of clusters which size $\xi_2$ decreases while becoming denser ($d_{2} \nearrow$) and less interpenetrated ($\xi_2/\xi_1 \searrow$).
As the flow cessation gels are obtained from the same carbon black dispersion, the density of carbon black particles should be conserved. There are two ways to calculate this density. Based on the carbon black volume fraction, the average density is $\bar \rho =\phi / a^3\simeq 900~\mu$m$^{-3}$. Based on the SAXS data, assuming an homogeneous network, we can measure the local particle density $\rho$. The unit cell of the network is defined by the structural length $\xi_1$. This unit cell of volume $\xi_1^3$ contains one cluster with $\left(\xi_2/a\right)^{d_2}$ particles so that $\rho=\left(\xi_2/a\right)^{d_2}/\xi_1^3$. In Fig. \ref{fig:Parameters}a, we measure $\rho\simeq7200$~$\mu$m$^{-3}$ a value larger than $\bar \rho$.
}

{\color{black} In Fig. \ref{fig:Parameters}b, we display the fit results based on the clusters of clusters model. As $\dot{\gamma}_0$ increases, we observe that the gel is composed of supperaggregates which become smaller ($\xi_2 \searrow$) and denser ($d_{2} \nearrow$) while the clusters that compose them become larger ($\xi_1 \nearrow$) and looser  ($d_{1} \searrow$).
Again we can compare $\rho$ and $\bar \rho$. The unit cell of the network is defined by the structural length $\xi_2$. In the volume $\xi_2^3$, there is exactly one superaggregates composed of $\left(\xi_2/\xi_1\right)^{d_2}$ clusters themselves composed of $\left(\xi_1/a\right)^{d_1}$ particles. Hence, we get $\rho=\left(\xi_2/\xi_1\right)^{d_2}\left(\xi_1/a\right)^{d_1}/\xi_2^3$. We measure $\rho\simeq2500$~$\mu$m$^{-3}$ a value again larger than $\bar \rho$.
}

Both approach verify that $\rho$ remains constant throughout the $\dot{\gamma}_0$ series (Fig.~\ref{fig:Parameters}). This infers that our approach is self-consistent. However, the value of $\rho$ calculated with the structure parameters is way larger compared to the one calculated based on the volume fraction of the carbon black particles, $\bar \rho \simeq 900~\mu$m$^{-3}$. This discrepancy between the two ways of calculating the density suggests that the gel network is heterogeneous. We dismiss shear banding as discussed in~\cite{Dages2022} and fractures which would lead to a strong enhanced scattering and a Porod regime at low $q$~\cite{Dages2021}. We vouch for a fractal network. This remains hypothetical as the hallmark of this fractal network is not measurable in the SAXS experiments. Indeed, this network fractal dimension should appear in $I(q)$ in the very low $q$-regime for $q_2<2\pi/\xi_2\sim 0.002$~nm$^{-1}$. As the lowest $q$ in the SAXS experiment is 0.001~nm$^{-1}$, we cannot probe the existence of $D$ nor measure its value.

 When the network is not homogeneous, $\rho > \bar \rho$. Assuming the gel network is fractal, is there a way to determine $D$ from $\bar \rho$  and $\rho$? This would greatly help as the measure of $D$ is not accessible in the SAXS experiment.
To tackle the issue of the local density of particles, Ref.\cite{Tang2008,Marangoni2021} used an heterogeneous mass distribution which has an impact on the rheological properties as derived by \cite{Shih1990,Wu2001,Mellema2002}. More precisely, the authors used the probablity of an inter-cluster bond at any location in a cluster chain to become stress-carrying given the condition that all other bonds are not stress-carrying as a constant. This assumption gives rise to an exponential probability distribution in terms of the number of inter-cluster bonds in a small unit of volume. This approach reads a final expression of the storage modulus as a power law of the particle volume fraction multiplied by $\left(1-\exp\left(-k\phi^b\right)\right)^\beta$ with $\left(k,b,\beta\right)$ some constants of the material. In the limit of low volume fraction, the usual power law in $\phi$ is recovered but the multiplicative factor allow a deviation from the power law at higher volume fraction. Even if we do not consider this approach for the rheological data, the perspective of a heterogeneous distribution of particles as in a fractal structure is a promising way to account for structural and rheological discrepancies. It is expected then that the local density $\rho$ can be related to the average density $\bar{\rho}$ through the influence of the fractal dimension but it remains a challenge out of the scope of this paper.

\section{Model of the gels elasticity}
\label{sec:rheo}
In this section, we first describe the common basis to derive the elastic modulus from the clusters of clusters $\phi$-power law model and the interpenetration of clusters $\phi$-power law model. We then derive an analytical expression of the elastic modulus for each model. Finally, using this analytical expressions together with the structure parameters displayed in Fig.~\ref{fig:Parameters}, we fit the rheology data presented in Fig.~\ref{fig:Intensity}(a) and discuss the rheology fit parameters. 

\subsection{Common basis for both models}
Both the clusters of clusters $\phi$-power law model and the interpenetration of clusters $\phi$-power law model belong to the class of $\phi$-power law models which aim to provide an analytical expression to fit the gel storage modulus $G'_{\infty}$ as shown in Fig.~\ref{fig:Intensity}(a). The derivation of the such models follow a top down approach. The macroscopic storage modulus $G'$ is \emph{a priori} related to the macroscopic stiffness $K$ of the system of size $L$~\cite{Mellema2002} by:
\begin{equation}
    G'=\frac{K}{L^{1+2\epsilon\alpha}}\label{eq:MacroElasticityCluster}.
\end{equation}
The nature of $K$ is solely defined by dimension analysis. Depending on the value of the exponent of $L$, $K$ is a stretching stiffness ($L^1$) or a bending stiffness ($L^3$). The exponent $\epsilon$~\cite{Mellema2002} allows one to have a mix of both stiffness as it varies from $\epsilon=0$ (pure stretching) to $\epsilon=1$ (pure bending). The exponent $\alpha$ indicates the proportion of the weak-link to strong-link regime~\cite{Wu2001}. In the weak-link limit where the inside of the clusters is stiffer than the connection between the clusters, $\alpha=0$.
In the strong-link limit where the inside of the cluster is less stiff than the connection between the clusters, $\alpha=1$.
The origin of the previous expression is purely dimensional trying to relate the macroscopic stiffness to the macroscopic storage modulus with precisely defined dimensions. The previous exponents are coming from mixed approached taking into account multiple types of phenomena. The extreme values of the exponents isolate the different cases depending on the type of interaction between the clusters (bending, stretching, dominated by the links, dominated by the internal stiffness of the clusters). 

The expression of $K$ is model dependant and is conditioned at the macroscopic scale by the network topology, at intermediate scale by the nature of the aggregates and their interactions and at the particle scale by the colloid-colloidal interactions as shown by Fig. \ref{fig:SketchMix}. Indeed, the gel elasticity can be decomposed \emph{a posteriori} in~
\cite{Shih1990,Wu2001,Mellema2002,Bouthier2022b} 
\begin{equation}
    G'=G_{\mathrm{cc}}S_\mathrm{agg}S_\mathrm{net}S_\mathrm{MBS}
    \label{eq:Decomposition}
\end{equation}
where $G_{\mathrm{cc}}$ is the elasticity resulting from the colloid-colloid interaction potential $\mathcal{U}$, $S_\mathrm{agg}$ is the scaling coming from the aggregation scale, $S_\mathrm{net}$ is the scaling contribution from the network scale and $S_\mathrm{MBS}$ is the scaling coming from the macroscopic influence of bending or stretching. This model is hierarchical.  The different factors come from the iterative demonstration procedure and the identification of the different contribution in the final formulas. The construction of the storage modulus directly expressing the different factors is impossible and needs the following demonstrations. However, we will emphasize these contributions in the final formulas of each model. The radius of the colloid $a$ sets the colloidal scale, the structural dimension $\xi_s$ set the scale at the aggregation level and serves as the coarse grain unit to construct the network. 
In our case, compared to the clusters $\phi$-power law models~\cite{Shih1990,Wu2001,Mellema2002}, we have  kept the colloidal scale unchanged and modified the aggregation and the network scale.

At the colloidal scale, the elasticity $G_{\mathrm{cc}}$ is set by the characteristics of the attractive interaction potential $\mathcal{U}$, namely its depth $U$ and  its range $\delta$. This leads, in the case of monodisperse colloidal spheres, to a storage modulus~\cite{Shih1990,Wu2001,Bouthier2022b}
\begin{equation}
    G_{\mathrm{cc}}=\frac{U}{\delta^2a}.
    \label{eq:Gcc}
\end{equation}
Typically, $U$ is few tens of $k_BT$, with $k_B$ the Boltzman constant and $T$ the temperature, and $\delta$ a few percent of the colloid radius $a$. $G_{\mathrm{cc}}$ is the only term in Eq.~(\ref{eq:Decomposition}) that has the dimensionnality of a shear modulus and therefore it sets the amplitude scale of the gel elasticity $G'$.

At the macroscopic scale, the network topology is determinant and two options are possible. Either the aggregates of structural length $\xi_s$ fill the space in a close packing configuration or they form a fractal structure of upper limit the size of the sample $L$ and a fractal dimension $D$. 
In the case of the fractal network, it is possible to distinguish the weak-link regime and the strong-link regime\cite{Shih1990,Wu2001}. On the one hand, in the weak-link regime, the connections between the clusters are weaker than the clusters themselves thus they dominate at the larger scale due to the assumption of springs in series. Therefore, the contribution of the network is 
\begin{equation}
    S_\mathrm{net}=1.
    \label{eq:S1}
\end{equation}
On the other hand, in the strong-link regime, the inside clusters cohesion is weaker than the links between clusters, hence it dominates at the larger scale. It is then relevant to introduce the dimension $\mathfrak{D}$ ({\color{black} also commonly named chemical dimension}) of the elastic back-bone of the network which is typically in the range $\left[1,1.5\right]$~\cite{Grassberger1985,Grassberger1992a,Grassberger1992b} and accounts for the network path that sustain the network stress. The contribution of the network is then 
\begin{equation}
    S_\mathrm{net}=\left(\frac{\xi_s}{L}\right)^\mathfrak{D}.
        \label{eq:S2}
\end{equation}
Both previous Eqs.~(\ref{eq:S1}) and (\ref{eq:S2}) may be unified,
\begin{equation}
    S_\mathrm{net}=\left(\frac{\xi_s}{L}\right)^{\mathfrak{D}\alpha}.
\end{equation}
The exponent $\alpha$ then indicates the weak-link and strong-link proportion, $\alpha\in\left[0,1\right]$\cite{Wu2001}.

In the close packing case, rather than adopting a random close packing structure, one assume a cubic crystal structure to simplify calculations~\cite{Shih1990,Mellema2002,Wu2001}. The local elastic entities of the network are organised in a crystal structure with springs in parallel and in series. Assuming that all distances are equal to $\xi_s$, there are $\left(L/\xi_s\right)^{\dim-1}$ springs in parallel, with $\dim$ the dimension of the euclidean space, themselves composed by $L/\xi_s$ springs in series. Summing the stiffnesses or the inverse of the stiffnesses for the springs in parallel or in series respectively, this reads to the factor
\begin{equation}
    S_\mathrm{net}=\left(\frac{L}{\xi_s}\right)^{\dim-2}.
\end{equation}
Finally, the macroscopic influence of bending or stretching $S_\mathrm{MBS}$ is characterised by a dimensionless factor to the exponent $1+2\epsilon\alpha$ as expressed in Eq.~(\ref{eq:MacroElasticityCluster}). This factor will be expressed thoroughly in the following sections of the paper. 

For reference, in the clusters $\phi$-power law model~\cite{Shih1990}, hypothesising an homogeneous network, the gel elasticity is
\begin{equation}
    \frac{G'}{G_{\mathrm{cc}}}=\begin{cases}
\left(\frac{\xi_{s}}{a}\right)^{2-\dim}=\phi^{\frac{\dim-2}{\dim-d}} & \text{weak-link regime}\\
\left(\frac{\xi_{s}}{a}\right)^{-\mathfrak{d}-\dim}=\phi^{\frac{\dim+\mathfrak{d}}{\dim-d}} & \text{strong-link regime}
\label{eq:shi}
\end{cases}
\end{equation}
In this model the structural length is the cluster size $\xi_s=\xi$ and $\phi$ is the particle volume fraction. $d$ and $\mathfrak{d}$ represent respectively the fractal dimension of the cluster and the dimension of the cluster back-bone.

\subsection{Clusters of clusters $\phi$-power law model}

In the clusters of clusters $\phi$-power law model~\cite{Bouthier2022b}, we consider that colloidal particles of radius $a$ assemble into clusters of  dimension $\xi_1$ and fractal dimension $d_{1}$ which themselves aggregates into  super-aggregates of  dimension $\xi_2$ and fractal dimension $d_{2}$ to form the gel network as sketched in the Fig.~\ref{fig:SketchMix}.  
We note respectively $\mathfrak{d}_1$ and $\mathfrak{d}_2$ the dimension of elastic back-bone of cluster and super-aggregates~\footnote{For details about chemical dimension or dimension of elastic back-bone, \cite{Grassberger1985,Grassberger1992a,Grassberger1992b} give some useful insights}. 
At the aggregation scale, the structural length scale used to built the network is $\xi_s=\xi_2$. The macroscopic stiffness $K$ is related to the superaggregate local elasticity $K_\mathrm{sa}$ through the close packing assumption bringing 
\begin{equation}
    K=\left(\frac{L}{\xi_2}\right)^{\dim-2}K_\mathrm{sa}=S_\mathrm{net}K_\mathrm{sa}.
    \label{eq:ClusterClusterNetwork}
\end{equation}
Moreover, the superaggregates local elasticity $K_{\mathrm{sa}}$ is related to the cluster local elasticity $K_c$ through the elastic back-bone scaling of the flocs which reads 
\begin{equation}
    K_\mathrm{sa}=\left(\frac{\xi_1}{\xi_2}\right)^{\mathfrak{d}_2\alpha}K_c,
\end{equation}
with $\alpha\in\left[0,1\right]$ the proportion of weak-link regime ($\alpha=0$) and strong-link regime ($\alpha=1$) according to \cite{Wu2001}. Furthermore, the cluster local elasticity is related to the local particle elasticity $K_p$ with the elastic back-bone scaling of the cluster which brings 
\begin{equation}
    K_c=\left(\frac{a}{\xi_1}\right)^{\mathfrak{d}_1\alpha}a^{2\epsilon\alpha}K_p.
\end{equation}
Hence, the aggregation level contribution is 
\begin{equation}
    S_\mathrm{agg}=\left(\frac{\xi_1}{\xi_2}\right)^{\mathfrak{d}_2\alpha}\left(\frac{a}{\xi_1}\right)^{\mathfrak{d}_1\alpha}.
\end{equation}
Finally, the local particle elasticity $K_p$ is expressed by 
\begin{equation}
    K_p=\frac{U}{\delta^2}.
\end{equation}
All the geometrical quantities may be related to the particle volume fraction through 
\begin{equation}
    \phi=\left(\frac{\xi_2}{L}\right)^{3-\dim}\left(\frac{\xi_1}{\xi_2}\right)^{3-d_2}\left(\frac{a}{\xi_1}\right)^{3-d_1}.\label{eq:ConservationMassCluster}
\end{equation}
Putting Eqs. (\ref{eq:MacroElasticityCluster})-(\ref{eq:ConservationMassCluster}) together reads
\begin{multline}
G'=\frac{U}{a\delta^{2}}\phi^{1+\frac{2\epsilon\alpha}{3-\dim}}\times\\
\left(\frac{\xi_{1}}{\xi_{2}}\right)^{\mathfrak{d}_2\alpha-2+d_2+\frac{d_2-\dim}{3-\dim}2\epsilon\alpha}\times\\
\left(\frac{a}{\xi_{1}}\right)^{\mathfrak{d}_1\alpha-2+d_1+\frac{d_1-\dim}{3-\dim}2\epsilon\alpha}
    \label{eq:ResultClusterCluster}
\end{multline}
and can be rewritten $ G'/G_\mathrm{cc}  =S_\mathrm{agg}S_\mathrm{net}S_\mathrm{MBS}$ with

\begin{equation}
\begin{array}{l}
\left\{
\begin{array}{ll}
S_\mathrm{net} & = \left(\phi^{-\frac{1}{3-\dim}}\left(\frac{\xi_{1}}{\xi_{2}}\right)^{\frac{3-d_{2}}{3-\dim}}\left(\frac{a}{\xi_{1}}\right)^{\frac{3-d_{1}}{3-\dim}}\right)^{\dim-2} \\
 S_{\mathrm{MBS}} & = \left(\phi^{-\frac{1}{3-\dim}}\left(\frac{\xi_{1}}{\xi_{2}}\right)^{\frac{6-d_{2}-\dim}{3-\dim}}\left(\frac{a}{\xi_{1}}\right)^{\frac{6-d_{1}-\dim}{3-\dim}}\right)^{-1-2\epsilon\alpha} \\
 S_{\mathrm{agg}} & = \left(\frac{\xi_{1}}{\xi_{2}}\right)^{\mathfrak{d}_{2}\alpha}\left(\frac{a}{\xi_{1}}\right)^{\mathfrak{d}_{1}\alpha}
\end{array}
\right.
\end{array}.
   \label{eq:ContributionClusterCluster}
\end{equation}

When looking at Eqs. (\ref{eq:ResultClusterCluster}) and (\ref{eq:ContributionClusterCluster}), there are some variables where the exponent has a denominator equals to $3-\dim$. This difficulty is actually present in most of the demonstrations in the literature \cite{Shih1990,Wu2001,Mellema2002} but is omitted considering proportional relationship between the relevant variables, the particle volume fraction and the microstructure variables. For example, in Eq. (10) in \cite{Shih1990}, there is the influence of the macroscopic size $L$ in the macroscopic stiffness and the particle volume fraction, the cluster size and the macroscopic size are related between each other. Hence, we know the existence of this difficulty which is consistent with the approach described usually in the literature. 

To avoid this difficulty, one can decide to impose $\epsilon=0$ to be in the stretching regime and to get 
\begin{equation}
    G'=\frac{U}{a\delta^{2}}\phi\left(\frac{\xi_{1}}{\xi_{2}}\right)^{\mathfrak{d}_2\alpha-2+d_2}\left(\frac{a}{\xi_{1}}\right)^{\mathfrak{d}_1\alpha-2+d_1},\label{eq:ModulusClusterClosePacking}
\end{equation}
which can be simplified in the weak-link regime ($\alpha=0$) to 
\begin{equation}
    G'=\frac{U}{a\delta^{2}}\phi\left(\frac{\xi_{1}}{\xi_{2}}\right)^{d_2-2}\left(\frac{a}{\xi_{1}}\right)^{d_1-2}.
\end{equation}

Another possibility of the model is to change the homogeneous network into heterogeneous fractal network. In this case, Eqs. (\ref{eq:ClusterClusterNetwork}) and (\ref{eq:ConservationMassCluster}) are replaced respectively by 
\begin{align}
    S_\mathrm{net}&=\left(\frac{\xi_2}{L}\right)^{\mathfrak{D}\alpha}\\
    \phi&=\left(\frac{\xi_{2}}{L}\right)^{3-D}\left(\frac{\xi_{1}}{\xi_{2}}\right)^{3-d_2}\left(\frac{a}{\xi_{1}}\right)^{3-d_1}
\end{align}
with $D$ the network fractal dimension and $\mathfrak{D}$ the dimension of the elastic back-bone. Eq. (\ref{eq:ResultClusterCluster}) is then replaced by

\begin{multline}
    G'= \frac{U}{a\delta^{2}}\phi^{\frac{\mathfrak{D}\alpha+1+2\epsilon\alpha}{3-D}} \times\\
    \left(\frac{\xi_{2}}{\xi_{1}}\right)^{\left(3-d_{2}\right)\frac{\mathfrak{D}\alpha+1+2\epsilon\alpha}{3-D}-\mathfrak{d}_{2}\alpha-1-2\epsilon\alpha}\times\\
    \left(\frac{\xi_{1}}{a}\right)^{\left(3-d_{1}\right)\frac{\mathfrak{D}\alpha+1+2\epsilon\alpha}{3-D}-\mathfrak{d}_{1}\alpha-1-2\epsilon\alpha}
    \label{eq:ModulusClusterFractal}
\end{multline}

which can be rewritten $G'/G_\mathrm{cc}  =S_\mathrm{agg}S_\mathrm{net}S_\mathrm{MBS}$ with
\begin{equation}
\begin{array}{l}
\left\{
\begin{array}{ll}
S_\mathrm{net} & =\left(\phi\left(\frac{\xi_{2}}{\xi_{1}}\right)^{3-d_{2}}\left(\frac{\xi_{1}}{a}\right)^{3-d_{1}}\right)^{\frac{\mathfrak{D}\alpha}{3-D}} \\
 S_{\mathrm{MBS}} & =\left(\left(\phi\left(\frac{\xi_{2}}{\xi_{1}}\right)^{3-d_{2}}\left(\frac{\xi_{1}}{a}\right)^{3-d_{1}}\right)^{\frac{1}{3-D}}\left(\frac{\xi_{2}}{\xi_{1}}\right)^{-1}\left(\frac{\xi_{1}}{a}\right)^{-1}\right)^{1+2\epsilon\alpha} \\
 S_{\mathrm{agg}} & =\left(\frac{\xi_{2}}{\xi_{1}}\right)^{-\mathfrak{d}_{2}\alpha}\left(\frac{\xi_{1}}{a}\right)^{-\mathfrak{d}_{1}\alpha}
\end{array}
\right.
\end{array}.
    \label{eq:ModulusClusterFractal_decomposition}
\end{equation}

\subsection{Interpenetration of clusters $\phi$-power law model}

In the Interpenetration of clusters $\phi$-power law model, the size $a$ still corresponds to the particle size but the size $\xi_1$ corresponds to the center-to-center distance between the clusters composed by particles of size $a$ and the size $\xi_2$ corresponds to the cluster size and fractal dimension $d_2$, as sketched on Fig.~\ref{fig:SketchMix}. The structure length scale $\xi_s$ for the network is now $\xi_s=\xi_1$. The macroscopic gel linear storage modulus $G'$ is again given by Eq.~(\ref{eq:MacroElasticityCluster}). To simplify the calculations, we dismissed the bending contributions so that $K$ is a purely stretching linear elastic stiffness ($\epsilon=0$). Adopting a close packing configuration for the network leads to
\begin{equation}
    S_\mathrm{net}=\left(\frac{L}{\xi_1}\right)^{\dim-2}.
    \label{eq:Snet}
\end{equation}
The mass conservation can then be written
\begin{equation}
    \phi=\left(\frac{a}{\xi_{1}}\right)^{3}\left(\frac{\xi_{2}}{a}\right)^{d_{2}}\left(\frac{\xi_{1}}{L}\right)^{3-\dim}.\label{eq:ConservationMass}
\end{equation}
For the aggregation level, we decompose the elementary effective stiffness of the clusters as in \cite{Wu2001} with 
\begin{equation}
    \frac{1}{K_\mathrm{eff}}=\frac{1}{K_c}+\frac{1}{K_\mathrm{ext}}+\frac{1}{K_i}
\label{eq:EffectiveStiffness}
\end{equation}
with $K_c$, $K_\mathrm{ext}$ and $K_i$ the elastic stiffness related to the inside of the cluster, the intermicroscopic (see \cite{Wu2001}) and the interpenetration of the cluster, respectively.

Let us now quantify $K_i$. $K_i$ is assumed to be proportional to the number of contact $N_i$ between interpenetrating clusters: $K_i=N_i k_i$, where $k_i$ is a reference interpenetration stiffness. The last expression comes from the fact we assumed that the contacts in the interpenetration zone are parallel springs: this justifies the additivity of the stiffnesses. We assume that the reference interpenetration stiffness $k_i$ is directly related to the depth of the interaction potential $U$ and the distance of interaction $\delta$ through $k_i=U/\delta^2$. Due to the fractal nature of the clusters, there are $N_i=\left(6V_i/\pi a^3\right)^\frac{d_2}{3}$ particles inside the intersection volume $V_i$ between two clusters. Geometrically assimilating clusters to spheres leads to an intersection volume \cite[p. 97]{Polyanin2007,Kern1967}

\begin{equation}
    V_i=\frac{\pi}{12}\xi_{2}^3\left(2+\frac{\xi_1}{\xi_2}\right)
    \left(1-\frac{\xi_1}{\xi_2}\right)^{2} \textbf{1}_{\left\{\xi_1<\xi_2\right\}}.
    \label{eq:vi}
\end{equation}
We assume that each particle brought by each cluster in $V_i$ form a contact adding rigidity to the whole system. Putting together the last expressions, we get

\begin{equation}
    K_i=\frac{U}{2\delta^2}\left(\frac{\xi_{2}}{a}\right)^{d_2}\left(1+\frac{\xi_1}{2\xi_2}\right)^{\frac{d_2}{3}}\left(1-\frac{\xi_1}{\xi_2}\right)^{\frac{2d_2}{3}}\textbf{1}_{\left\{\xi_1<\xi_2\right\}}.
    \label{eq:Model1}
\end{equation}

Let us now compare $K_i$ with $K_c$ and $K_\mathrm{ext}$. There are different ways to consider that $K_i\ll \min\left(K_c,K_\mathrm{ext}\right)$. To simplify the comparison, following \cite{Wu2001}, we write 
\begin{equation}
    \frac{1}{K_c}+\frac{1}{K_\mathrm{ext}}=\frac{1}{K_c}\left(\frac{K_c}{K_\mathrm{ext}}\right)^\alpha
\end{equation}
with $\alpha\in\left[0,1\right]$ allowing to make a continuous transition between the weak-link and the strong-link regime.
A first way to compare $K_i$ with $K_c\left(K_\mathrm{ext}/K_c\right)^\alpha$ is to say that the system is in the regime $\xi_2/\xi_1\gtrsim1$. Thus, one can re-write Eq.~(\ref{eq:Model1}) as

\begin{equation}
    K_i\underset {\xi_2/\xi_1\gtrsim1}{\propto} \frac{U}{2\delta^2} \left(1-\frac{\xi_1}{\xi_2}\right)^{\frac{2d_2}{3}}.
\end{equation}
$K_i$ depends strongly on the distance of $\xi_2/\xi_1$ from unity. Therefore, $K_i$ is negligible when $\xi_2/\xi_1\gtrsim1$ compared to $K_c\left(K_\mathrm{ext}/K_c\right)^\alpha$ and, in Eq.~(\ref{eq:EffectiveStiffness}), we get $K_\mathrm{eff}\approx K_i$.

The other way to consider the system is, following previous approaches in \cite{Kantor1984a,Kantor1984b,Shih1990,Wu2001,Mellema2002,Wessel1992}, estimating $K_c\left(K_\mathrm{ext}/K_c\right)^\alpha\propto\xi_1^{-\mu}$ with $\mu\in\left[1,5\right]$ function of the fractal dimension $D$, the dimension of the elastic back-bone and the regime of strong-link or weak-link because $\xi_1$ is similar to a cluster size with contact. Recalling that $\left(\xi_2/a\right)^{d_2}\propto\xi_1^3$, one gets in this case

\begin{equation}
    \frac{K_i}{K_c}\left(\frac{K_c}{K_\mathrm{ext}}\right)^\alpha\propto\xi_1^{3+\mu}\left(1+\frac{\xi_1}{2\xi_2}\right)^{\frac{d_2}{3}}\left(1-\frac{\xi_1}{\xi_2}\right)^{\frac{2d_2}{3}}.
\end{equation}

Assuming that $\xi_1$ does not vary much, $K_i/K_c\left(K_c/K_\mathrm{ext}\right)^\alpha$ is governed by the values of $\xi_2/\xi_1\mapsto\left(1+\xi_1/(2\xi_2)\right)^{\frac{d_2}{3}}\left(1-\xi_1/\xi_2\right)^{\frac{2d_2}{3}}$ on $\left[1.2,1.8\right]$. $K_i/K_c\left(K_c/K_\mathrm{ext}\right)^\alpha$ is, according to \citet{Dages2022}, between 0.1 and 0.3. Therefore, one can assume that $K_i\ll K_c\left(K_\mathrm{ext}/K_c\right)^\alpha$, at least for the first values, and following Eq.~(\ref{eq:EffectiveStiffness}), we get $K_\mathrm{eff}\approx K_i$.

Generally, as $K_i$ is getting closer to $K_c\left(K_\mathrm{ext}/K_c\right)^\alpha$, it becomes difficult to consider that only one phenomenon prevails. If one wants to completely understand the balance between the different contributions, one needs to model both phenomena and their coupling. This is not the goal of this model which tries to give some orders of magnitude without exhaustively modeling the system.

This final expression of the interpenetration $\phi$-power law model is then
\begin{equation}
    G'=\frac{U}{2a\delta^{2}}\left(1+\frac{\xi_{1}}{2\xi_{2}}\right)^{\frac{d_{2}}{3}}\left(1-\frac{\xi_{1}}{\xi_{2}}\right)^{\frac{2d_{2}}{3}}\phi\left(\frac{\xi_{1}}{a}\right)^{2}
\end{equation}
which can be rewritten $G'/G_\mathrm{cc}  =S_\mathrm{agg}S_\mathrm{net}S_\mathrm{MBS}$ with
\begin{equation}
\begin{array}{l}
\left\{
\begin{array}{ll}
S_\mathrm{net} & = \left(\phi\left(\frac{\xi_{1}}{a}\right)^{3}\left(\frac{\xi_{2}}{a}\right)^{-d_{2}}\right)^{\frac{2-\dim}{3-\dim}} \\
 S_{\mathrm{MBS}} & = \frac{a}{\xi_{1}}\left(\phi\left(\frac{\xi_{1}}{a}\right)^{3}\left(\frac{\xi_{2}}{a}\right)^{-d_{2}}\right)^{\frac{1}{3-\dim}} \\
 S_{\mathrm{agg}} & =\frac{1}{2}\left(\frac{\xi_{2}}{a}\right)^{d_{2}}\left(1+\frac{\xi_{1}}{2\xi_{2}}\right)^{\frac{d_{2}}{3}}\left(1-\frac{\xi_{1}}{\xi_{2}}\right)^{\frac{2d_{2}}{3}}
\end{array}
\right.
\end{array}.
\label{eq:ModulusInterp}
\end{equation}

The dimension of the network $\dim$ is not required in the final expression due to the contribution of the effective volume fraction through the particle volume fraction and the fractal dimension related to $\xi_1$.

If we now replace the homogeneous network by a heterogeneous network of fractal dimension $D$ and {\color{black} back-bone} dimension $\mathfrak{D}$, Eqs.~(\ref{eq:Snet}) and (\ref{eq:ConservationMass}) become respectively
\begin{align}
    S_\mathrm{net}&=\left(\frac{\xi_{1}}{L}\right)^{\mathfrak{D}\alpha}\\
    \phi&=\left(\frac{a}{\xi_{1}}\right)^{3}\left(\frac{\xi_{2}}{a}\right)^{d_{2}}\left(\frac{\xi_{1}}{L}\right)^{3-D}.
\end{align}
The storage modulus is then 
\begin{multline}
    G'=\phi^{\frac{1+\mathfrak{D}\alpha}{3-D}}\frac{U}{2a\delta^{2}}\left(\frac{\xi_{1}}{a}\right)^{\frac{3\mathfrak{D}\alpha+D}{3-D}}\left(\frac{\xi_{2}}{a}\right)^{d_{2}\frac{2-D-\mathfrak{D}\alpha}{3-D}}\times\\
    \left(1+\frac{\xi_{1}}{2\xi_{2}}\right)^{\frac{d_{2}}{3}}\left(1-\frac{\xi_{1}}{\xi_{2}}\right)^{\frac{2d_{2}}{3}}
    \label{eq:ModulusInterp2}
\end{multline}
which can be rewritten $G'/G_\mathrm{cc}  =S_\mathrm{agg}S_\mathrm{net}S_\mathrm{MBS}$ with
\begin{equation}
\begin{array}{l}
\left\{
\begin{array}{ll}
S_\mathrm{net} & = \left(\phi\left(\frac{\xi_{1}}{a}\right)^{3}\left(\frac{\xi_{2}}{a}\right)^{-d_{2}}\right)^{\frac{\mathfrak{D}\alpha}{3-D}} \\
 S_{\mathrm{MBS}} & =\left(\frac{\xi_{1}}{a}\right)^{-1}\left(\phi\left(\frac{\xi_{1}}{a}\right)^{3}\left(\frac{\xi_{2}}{a}\right)^{-d_{2}}\right)^{\frac{1}{3-D}} \\
 S_{\mathrm{agg}} & =\frac{1}{2}\left(\frac{\xi_{2}}{a}\right)^{d_{2}}\left(1+\frac{\xi_{1}}{2\xi_{2}}\right)^{\frac{d_{2}}{3}}\left(1-\frac{\xi_{1}}{\xi_{2}}\right)^{\frac{2d_{2}}{3}}
\end{array}
\right.
\end{array}.
\label{eq:ModulusInterp2b}
\end{equation}

In conclusion of this section \ref{sec:Model}, contrary to the classical $\phi-$power law model in Eq.~(\ref{eq:shi}), we observe that the scaling of the gel elasticity not only depends on $\phi$ but also the ratios between the length scales of the gel.

\subsection{Discussion of the gel elasticity}
\label{sec:Experiment}

Next, we turn to the elastic properties of the carbon back gels obtained through the flow cessation protocol. Fig.~\ref{fig:Intensity} displays the evolution of the elastic modulus $G'_\infty$ of the gel measured during a frequency sweep experiment in the low frequencies domain as a function of $\dot{\gamma}_0$. The gel becomes weaker as $\dot{\gamma}_0$ increases.

\begin{figure}
    \centering
    \includegraphics[width=\columnwidth]{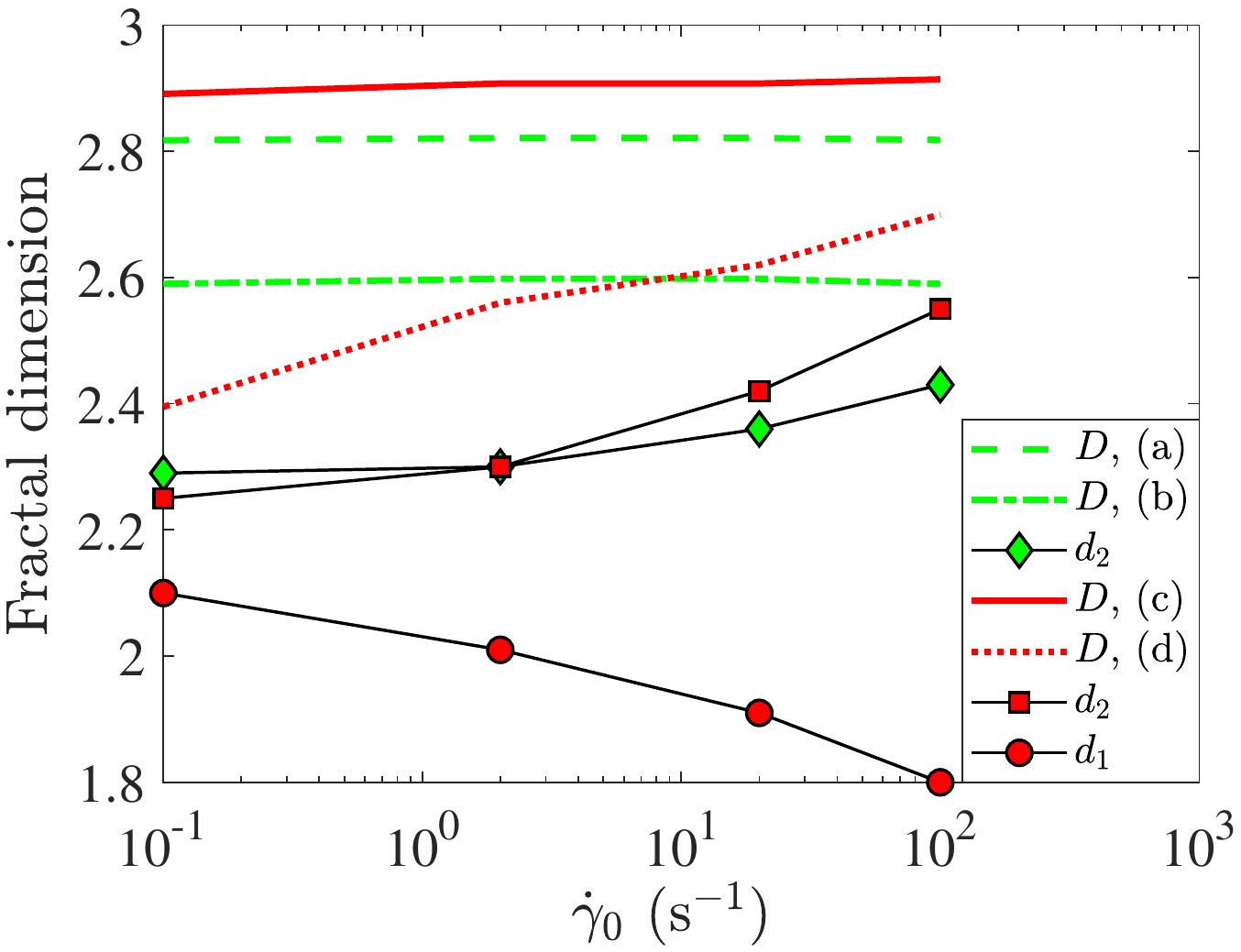}
    \caption{Evolution of the gel network fractal dimension $D$ as a function of $\dot{\gamma}_0$ for the models (a) through (d) which parameters are listed in Tab.~\ref{tab:model}. The fractal dimension $d_1$ and $d_2$ from Fig.~\ref{fig:Parameters} are also plotted for comparison. The green data is related to the interpenetrating cluster models whereas the red data is related to the cluster of clusters model.
    }
    \label{fig:fractal}
\end{figure}

Fitting $G'_\infty$ with the $\phi$-power law models is complex given the number of parameters in the equations. The fit is underdetermined: there are more parameters than available data to fit. To remain relevant, we constrain the fit parameters. First, we set $\xi_1$, $d_1$, $\xi_2$ and $d_2$ to the values of the SAXS models as shown in Fig.~\ref{fig:Parameters}. Second, we set the elasticity scale $G_\mathrm{cc}$ and take the value from the simulations by  Varga et al.~\cite{Varga2019}: $G_\mathrm{cc} = 6 \times 10^7$~Pa with $U= 30$~$k_BT$ and $\delta = 0.01a$. Third, based on the SAXS analysis, we only consider the models with a fractal gel network. This hypothesis is indeed verified as all models in the homogeneous network limit fail to fit $G'_\infty$ with the constrain $G_\mathrm{cc} = 6 \times 10^7$~Pa.
Fourth, we choose to constrain the fit to explore the parameter space and decide to study limiting cases: the weak-link limit $\alpha=0$ (model (a) and (c)) and the strong-link limit $\alpha=1$ in the stretching limit $\epsilon=0$ (model (b) and (d)) or the bending limit $\epsilon=1$ (model (e)). Again to simplify the model, when $\alpha \neq 0$, we constrained the dimensions of the elastic back-bones to $\mathfrak{D}=\mathfrak{d}_1=\mathfrak{d}_2=1.25$, an intermediate value between its extremes 1 and 1.5~\cite{Grassberger1985,Grassberger1992a,Grassberger1992b}.  Finally, we note, that, if $D \simeq d_2$, there is no contrast in the scattering intensity $I(q)$ between the aggregate of fractal dimension $d_2$ and the gel network of fractal dimension $D$. Therefore, one must have $d_2 \neq D$.
With those constrains, the only fitting parameter is $D$. 
Model (e) fails to fit the data whereas models (a) to (d) successfully fit $G_{\infty}$ as shown in Fig.~\ref{fig:Intensity}. The fit parameters are listed in Tab.~\ref{tab:model} and $D$ as a function of $\dot{\gamma}_0$ is plotted in Fig.~\ref{fig:fractal}.

\begin{table}
\centering
\begin{tabular}{|c || c| c | c || c |c ||c  |}
 \hline
Model & \textbf{$\alpha$} & $\epsilon$ & $G_\mathrm{cc}$ (Pa)  & $\mathfrak{d}_1 = \mathfrak{d}_2 = \mathfrak{D}$ & $D$ & Fit? \\ 
 \hline
 \multicolumn{7}{|c|}{} \\ 
 \multicolumn{7}{|c|}{ \textbf{Interpenetration of clusters $\phi$-power law model}, Eq.~(\ref{eq:ModulusInterp2})} \\
   \multicolumn{7}{|c|}{Parameters $\xi_1$, $\xi_2$ and $d_2$ are set by the experiments, Fig.~\ref{fig:Parameters}(a) } \\
 \hline
 (a) & 0 & - & 6 $\times$ 10$^{7}$ & - & 2.84 & yes      \\ 
  \hline
 (b) & 1 & - & 6 $\times$ 10$^{7}$ & 1.25 & 2.64 & yes      \\ 
   \hline
  \multicolumn{7}{|c|}{} \\ 
 \multicolumn{7}{|c|}{ \textbf{Clusters of clusters $\phi$-power law model}, Eq.~(\ref{eq:ModulusClusterFractal}) } \\
  \multicolumn{7}{|c|}{Parameters $\xi_1$, $d_1$, $\xi_2$ and $d_2$ are set by the experiments, Fig.~\ref{fig:Parameters}(b) } \\
 \hline

 (c) & 0 & - & 6 $\times$ 10$^{7}$ & -  & 2.91 & yes   \\
  \hline
 (d) & 1 & 0 & 6 $\times$ 10$^{7}$ & 1.25  & Fig.~\ref{fig:fractal} & yes  \\
  \hline
 (e) & 1 & 1 & 6 $\times$ 10$^{7}$ & 1.25 & - & no   \\ 
   \hline  
\end{tabular}
\caption{ Parameters of the models that fit $G'_{\infty}$ in Fig.~\ref{fig:Intensity}. $\alpha$ and $\epsilon$ define the models. We set $G_\mathrm{cc} = 6 \times 10^7$~Pa as in \cite{Varga2019}. $\xi_1$, $d_1$, $\xi_2$ and $d_2$ are determined by the experiments (Fig.~\ref{fig:Parameters}). $\mathfrak{D}$, $\mathfrak{d}_1$, $\mathfrak{d}_2$ are kept constant to 1.25.  $D$ is the free fit parameter.  Its values as a function of $\dot{\gamma}_0$ are displayed in Fig.~\ref{fig:fractal}. Model (e) could not fit the data for any $D$ value.
}
    \label{tab:model}
\end{table}

The models (a) through (d) are not discriminatory. In the weak link limit, models (a) and (c) yield a very high value of $D$, and the gel network is almost homogeneous. In the strong link limit, models (b) and (d) yield a weaker value of $D$, around 2.6. In all cases, the models show a hierarchical organisation of the fractal dimensions, $D>d_2>d_1$ or $D>d_2$ if $d_1$ does not exist as in the cluster interpenetration model. It is known~\cite{zaccone2009b,conchuir2014,Ruan2020,Varga2019,Jamali2019a,Jacquin2018,Jamali2020,Jamali2019b} that clusters under shear tend to become denser. A usual understanding of the phenomenon is that the competition between attractive forces due to the interaction potential and disaggregating forces due to shear impose the clusters size and fractal dimension. It is usually assumed that the cluster grow with decreasing shear rate and it is assumed but rarely measured that the cluster fractal dimension is constant \cite{Wessel1992,Kimbonguila2014,Eggersdorfer2010}. In turbulent flows of non-Brownian particles it is consistently measured that the clusters become larger and looser as the shear rate decreases~\cite{spicer1998, bubakova2013}.

However there is no clear understanding of this trend.
Measurements presented in Fig.~\ref{fig:Intensity} shows that, in the case of carbon black dispersions, the picture is much more complex and involves hierarchical structures with varying fractal dimensions. 
Moreover, the clusters are not isolated but form a space spanning network. Assuming that the structure during flow is very similar to one after flow cessation, the question is then, how does the flow before flow cessation propagate from large to small length scales?

In turbulence~\cite{Kolmogorov1941}, the energy cascades from the large to the small scales. Loss in energy only occurs at the small scales due to viscous dissipation.
Hence, the largest structure in the system, here the network of dimension $L$ sees directly the imposed shear flow which sets its fractal dimension $D$. For the aggregation level, the shear flows perceived by the intermediate clusters is diminished by the surrounding higher levels organisation. The intermediate structures have then less constraints to build up their structure which allows lower fractal dimension or less dense structures. To give some quantitative arguments, in the turbulent energy cascade \cite{Kolmogorov1941}, the rate of dissipation of energy $e$, which is also the rate of energy brought by the larger scales to the smaller scales, is independent of the length scale and considered constant after being set up by the macroscopic flow. Assessing the values of the rate of dissipation in our system, there is no inertia and only purely viscous dissipation. Hence, the rate of energy dissipation per unit volume is 
\begin{equation}
    e=\frac{\eta}{2}\left\lVert\boldsymbol{\nabla u}\right\rVert^2\approx\frac{\eta u^2}{h^2}
\end{equation}
with $\eta$ the dynamic viscosity, $u$ the velocity and $h$ a certain length scale over which the velocity vary. It is necessary to estimate the velocity $u$ and the length scale $h$. Using the Darcy law framework, the porosity of the gel is given by $1-\phi_\mathrm{eff}\left(l\right)$ with $\phi_\mathrm{eff}$ the effective volume fraction of particles-cluster-superaggregates for a certain window size $l$. The effective volume fraction $\phi_\mathrm{eff}$ increases with $l$, tends to $\phi$ when $l=0$, 1 when $l=L$, the network size. $\phi_\mathrm{eff}$ also depends on the particle volume fraction $\phi$, the different typical length scales $\xi_i$ and the fractal dimensions $d_i$ and $D$. However, the global trend of the function $\phi_\mathrm{eff}$ is sufficient. Assuming that a fixed flow rate per unit area $Q$ is fixed and in order to recover the fluid volume, the velocity and the length scale are given by $Q/u=\left(h/l\right)^3=1-\phi_\mathrm{eff}\left(l\right)$. Therefore, the rate of energy dissipation per unit volume is given by 
\begin{equation}
    e\approx\frac{\eta Q}{l^2}\left(1-\phi_\mathrm{eff}\left(l\right)\right)^{-\frac{8}{3}}.
\end{equation}
To conclude the demonstration, in our case $e$ decreases when $l$ decreases from $L$ to $a$ as opposed to the turbulent case where $e$ is constant. Thus, the energy is not continuously distributed among all the length scales from $L$ to $a$ but there is a decrease of the amount of energy brought to the smaller scales which ends up to decrease the constraints on the intermediate structures and allows for looser structures. If we pursue such reasoning it is possible to find then a decreasing density of the material, in terms of fractal dimension for example, from the largest scale to the lowest scale. This continuum briefly presented in \cite{Bouthier2022b} can find an experimental demonstration here.

This demonstration support two hypothesis, (i) the effective flow rate decreases as it propagates from large to small scale structures and (ii) the fractal dimension adopted by the structure decreases as the flow rate decreases. We therefore expect that the fractal dimension diminish as we go to lower length scales: $D>d_2>d_1$ as inferred by the data and the fits.

Following the fit results, the models (a-e) show that the carbon black gel elasticity come from stretching the bonds at all length scales and that bending can be dismissed ($\epsilon=0$). 
Bending is prevalent in gels forming strands such as arrested phase separation gels~\cite{gibaud2013} but also in fractal colloidal gels with a fractal dimension $d\sim 2$ where long thin chains of colloid come into play~\cite{pantina2006, dinsmore2006}. In the carbon black gels formed through flow cessation, its fractal structure is very dense, especially at large length scales, and seems to prevent bending from contributing significantly to the gel elasticity. 

The derivation of the three length scales rheological models opens future work directions. First, revisiting the classical $\phi$-power law model~\cite{Shih1990} which is widely used in the literature, we have pointed out an issue which has been eluded  up to now: the elasticity goes as $\phi^{\frac{1}{3-\dim}}$ and therefore diverges when the euclidean space dimension is equal to $\dim=3$ (Eq.~\ref{eq:ContributionClusterCluster}~\&~\ref{eq:ModulusInterp}). We have shown that this problem can be avoided when taking into account a fractal gel network (Eq.~\ref{eq:ModulusClusterFractal_decomposition}~\&~\ref{eq:ModulusInterp2b}) or having $\epsilon=0$ (Eq.~\ref{eq:ModulusClusterClosePacking}). Nevertheless this issue remains to be solved in future work. Second, we have developed an approach allowing to go continuously from the weak link to the strong link regime and from the stretching  to the bending limit. Those hypothesises based solely on scaling arguments (Eq.~\ref{eq:MacroElasticityCluster}) would benefit from in-depth numerical simulation or theoretical investigation. Finally, we hope that the case of three length scales colloidal gels is not specific to carbon black gels and that our model can be confronted in the future to other experimental systems.   

\section{Conclusion}

\label{sec:Conclusion}

We derived the mechanical properties of colloidal gels with a hierarchical structure composed of three distinct length scales.
In the cluster of cluster approach we considered that the particles of size $a$ aggregate in clusters of size $\xi_1$ and fractal dimension $d_1$ which themselves aggregate in superaggregates of size $\xi_2$ and fractal dimension $d_2$ to form the gel network of fractal dimension $D$. In the interpenetrating clusters approach we considered that the particles of size $a$ condensate into clusters which interpenetrate one-another to form the gel network.

Those three length scales rheological models are the main results of this paper. They provide an analytical expression of the gel elasticity incorporating a continuum description of the gel at the colloidal scale, the aggregation scale and the network scale. We have extended the historical $\phi$-power law model~\cite{Shih1990}, built to describe classical colloidal gels with two distinct length scales, in three directions.
First, at the aggregation scale we have developed the clusters of clusters and the interpenetration cluster approach. Second, we have worked on the gel network topology and added to the homogeneous network the possibility to form fractal networks which spans the system from the structuring aggregate size to the network size with a heterogeneous distribution of mass and stress-bearing chains. Third, we have provided an analytical expression of the elasticity that allows one to continuously transition from the weak-link to the strong-link regime (exponent $\alpha$) and from a stretching to a bending stiffness (exponent $\epsilon$).
Both models require structural measurements as an input. We suggest to measure the gel structure using SAXS and analyse the scattering intensity using the global scattering functions proposed by Beaucage.

Those rheological and structural models are then utilized to revisit experiments carried out on carbon black gels formed through flow cessations~\cite{Dages2022}. Results indicate that both approaches fit successfully in a self-consistent manner the experimental data: we could not discriminate between the clusters of clusters and the interpenetrating clusters approach to fit the carbon black data.
{\color{black} Indeed, to decide which model is more appropriate, we would need information on the structure of the gel network on length scales larger than the micron. Such information might be accessible using rheo-X-ray tomography~\cite{maire2014}.
Despite missing the gel network structure on length scale larger than a few microns, we were able to establish that the gel elasticity originate from stretching the bonds at all length scales and that bending can be dismissed.} We also showed, that for all $\dot{\gamma}_0$, there is a hierarchy in the fractal dimensions $D>d_2>d_1$ that follows the structural hierarchy $L>\xi_2>\xi_1>a$. We hypothesized that this hierarchy originate from the flow energy input before flow cessation which decreases continuously due to viscous dissipation when moving to smaller length scales, releasing the constraints on the intermediate structures and allowing looser and looser structures to form as we approach the colloidal scale.

\section*{Author Contributions}

L.V.B. designed the models. T.G. gathered and analysed the experimental data. L.V.B. and T.G. wrote the paper.

\acknowledgements
We thank the ESRF for beamtime at the beamline ID02 (proposal SC5099) and Theyencheri Narayanan and Lauren Matthews for the discussions and technical support for the USAXS measurements. This work was supported by the Région Auvergne-Rhône-Alpes ``Pack Ambition Recherche", the LABEX iMUST (ANR-10-LABX-0064) of Université de Lyon, within the program "Investissements d'Avenir" (ANR-11-IDEX-0007), the ANR grants (ANR-18-CE06-0013 and ANR-21-CE06-0020-01). This work benefited from meetings within the French working group GDR CNRS 2019 ``Solliciter LA Matière Molle" (SLAMM).


%

\end{document}